\documentclass[12pt]{article}
\usepackage{amsmath}
\usepackage{graphicx}
\usepackage{enumerate}
\usepackage{url} 
\usepackage{caption}
\usepackage{verbatim}
\usepackage{amssymb}
\usepackage{mathrsfs}
\usepackage{upgreek}
\usepackage{xcolor}
\usepackage{bm}
\usepackage{float}
\usepackage{nccmath}
\usepackage{mathtools}
\usepackage{natbib} 
\usepackage{xcolor}
\usepackage{algorithm}
\usepackage{algpseudocode}
\usepackage{hyperref}
\usepackage{setspace}
\usepackage{titlesec}
\usepackage{bbm}
\usepackage{soul} 

\newcommand{\nate}[1]{\textcolor{red}{\textsf{[#1]}}}

\newcommand{\bi}{\begin{itemize}}
\newcommand{\ei}{\end{itemize}}

\usepackage{xcolor}
\hypersetup{
    colorlinks,
    linkcolor={red!50!black},
    citecolor={blue!50!black},
    urlcolor={blue!80!black}
}

\newcommand{\blind}{1}

\begin{document}



\def\spacingset#1{\renewcommand{\baselinestretch}%
{#1}\small\normalsize} \spacingset{1}


\if1\blind
{
  \title{\bf Statistical Analysis of Quantitative Cancer Imaging Data }
  \author{Shariq Mohammed$^{1,2}$, Maria Masotti$^{3}$, Nathaniel Osher$^{3}$, \\ Satwik Acharyya$^{4}$, Veerabhadran Baladandayuthapani$^{3,5,}$\thanks{Corresponding author: \texttt{veerab@umich.edu}.}\\
\small{	$^{1}$Department of Biostatistics, Boston University, Boston, MA} \\
\small{	$^{2}$Rafik B. Hariri Institute for Computing and Computational Science \& Engineering, }\\ \small{Boston University, Boston, MA} \\
\small{	$^{3}$Department of Biostatistics, University of Michigan, Ann Arbor, MI} \\
\small{	$^{4}$Department of Biostatistics, University of Alabama at Birmingham, Birmingham, AL} \\
\small{	$^{5}$Department of Computational Medicine and Bioinformatics, University of Michigan, Ann Arbor, MI} \\}
  \maketitle
} \fi

\if0\blind
{
  \bigskip
  \bigskip
  \bigskip
  \begin{center}
    {\LARGE\bf Title}
\end{center}
  \medskip
} \fi


\bigskip
\begin{abstract}

Recent advances in types and extent of medical imaging technologies has led to proliferation of multimodal quantitative imaging data in cancer. Quantitative medical imaging data refer to numerical representations derived from medical imaging technologies, such as radiology and pathology imaging, that can be used to assess and quantify characteristics of diseases, especially cancer.  The use of such data in both clinical and research setting enables precise quantifications and analyses of tumor characteristics that can facilitate objective evaluation of disease progression, response to therapy, and prognosis. The scale and size of these imaging biomarkers is vast and presents several analytical and computational challenges that range from high-dimensionality to complex structural correlation patterns.  In this review article, we summarize some state-of-the-art statistical methods developed for quantitative medical imaging data ranging from topological, functional and shape data analyses to  spatial process models. We delve into common imaging biomarkers with a focus on radiology and pathology imaging in cancer, address the analytical questions and challenges they present, and highlight the innovative statistical and machine learning models that have been developed to answer relevant scientific and clinical questions. We also outline some emerging and open problems in this area for future explorations. 

\end{abstract}

\noindent%
{\it Keywords:}   Functional Data Analyses;  Pathology Imaging; Radiology Imaging; Spatial Process Modeling; Spatial Multiplex Imaging; Topological Data Analyses
\vfill

\newpage
\spacingset{1.9} 
\section{Introduction}
\label{sec:intro}


The rapid evolution of medical imaging technologies has been a cornerstone in the advancement of modern medicine, allowing the assessment of the complex interplay of biological processes that underpin the evolution and progression of disease. Notably, imaging has revolutionized oncology, where it plays a crucial role in diagnosing, staging, and monitoring disease and treatment response in cancer patients. As the digitization of medicine progresses, the breadth and depth of quantitative imaging data have expanded exponentially, providing an unprecedented opportunity to extract numerically precise, biomarker-based insights especially from radiology and pathology images \citep{oconnor2017}. Quantitative medical imaging data encapsulate numerical representations of the anatomical and functional characteristics of tissues, offering a vital source of information for the precise assessment and quantification of disease features \citep{sullivan2015}. For example, in oncology, the emergence of imaging biomarkers has ushered in a new paradigm for the objective evaluation of tumor characteristics, enabling detailed analyses of spatial distribution patterns of markers in the tumor microenvironment. These biomarkers are pivotal for non-invasive evaluation of disease states such as interpreting disease progression, gauging therapeutic response, and predicting prognosis with a level of detail that was previously unattainable \citep{gillies2016}. 


However, the complexity and volume of imaging data present significant analytical and computational challenges. The sheer scale and complexity of the data, characterized by high-dimensionality and intricate structural correlations, necessitate the development of sophisticated statistical methods capable of handling such data to distill actionable knowledge \citep{criminisi2013}. Rigorous statistical models and computational techniques have the potential to harness the full potential of imaging data, allowing clinicians and biomedical researchers to navigate the nuances of disease processes with greater confidence \citep{zhao2016}. 
These analytical and computational tools are not only refining our existing methodologies but are also paving the way for the discovery of novel biomarkers, thereby expanding the horizons of medical imaging.

In this article, we review some state-of-the-art statistical methods developed for the analysis of quantitative medical imaging data. We delve into the common imaging technologies and biomarkers with a focus on radiology and pathology imaging in cancer, address the analytical challenges they present, and highlight the innovative statistical models that have been recently developed to tackle these challenges. In Section 2, we discuss advanced statistical methods employing tools from topological and functional data analysis, as well as elastic shape analysis, with a focus on analyzing tumor morphology, texture and volume from radiology images. Section 3 delves into digital pathology, with a specific focus on the analysis of histopathological images. It emphasizes the significance of extracting meaningful features from tissue images using spatial models. These features not only offer insights into the intricate interactions within the tumor microenvironment and its clinical relevance but also serve as essential components in predictive modeling. Section 4 focuses on analyses of modern spatial multiplex cellular imaging data and elaborates on the computational and statistical methodologies tailored for handling such datasets. Finally, in Section 5, we outline some emerging and open problems in these areas.

\section{Radiology Imaging}
\label{sec:Radio_imaging}

\paragraph{Imaging technologies} Radiology imaging is a fundamental component of modern medical diagnosis and therapy. Common diagnostic imaging techniques used to visualize internal body structures and diagnose various medical conditions include X-rays, Computed Tomography (CT), Magnetic Resonance Imaging (MRI), and Positron Emission Tomography (PET) scans \citep{elangovan2016medical}. These imaging modalities play vital roles in modern medicine, enabling accurate diagnoses and guiding treatment decisions for various health conditions. X-ray images are generally used for detecting bone fractures, arthritis, and infections \citep{calabrese2014historical}, and CT scans to detect trauma injuries, tumors, vascular disease, and guide biopsies \citep{merino1976computerized}. MRI scans are useful for diagnosing aneurysms, stroke, tumors, and joint injuries \citep{berger2002does}, and PET scans are crucial in diagnosing cancer, heart disease, and Alzheimer's disease \citep{kapoor2020pet}. Functional magnetic resonance imaging (fMRI) is used to detect changes in blood flow \citep{buxton2013physics}, and is primarily focused on studying brain function rather than anatomical structure. While fMRI is related to radiology due to its use of MRI technology, it is often considered a separate field within neuroscience. 


\paragraph{Cancer and MRI} Cancer is a complex and multifaceted disease characterized by the uncontrolled growth and spread of abnormal cells within the body. There are numerous types of cancer, each classified based on the specific location of origin and the behavior of the abnormal cells. Common types include brain cancer, breast cancer, lung cancer, prostate cancer, colorectal cancer, and leukemia, among others \citep{neapolitan2015pan}. Each type of cancer presents unique challenges in terms of diagnosis, treatment, and prognosis, highlighting the importance of precise and accurate imaging techniques for early detection and management. Among the various imaging modalities used in cancer diagnosis and monitoring, MRI plays a crucial role as it is non-invasive and has ability to provide detailed and high-resolution images of internal structures, while also providing feasibility for repeated imaging studies \citep{yankeelov2011quantitative}. It is particularly valuable for its ability to detect and characterize tumors, assess tumor size and extent, evaluate the tumor region, nearby tissues and organs for potential metastasis, and monitor disease progression and treatment response over time \citep{schafer2013low}. In this review article, we will primarily focus on methods for the analysis of MRI scans in brain cancer while highlighting relevant statistical literature for analyzing MRI and CT data in other diseases. MRI provides multi-planar and multi-modal imaging capabilities, allowing clinicians to view the brain from various angles and perspectives, facilitating comprehensive tumor assessment and surgical planning \citep{zhang2017multimodal}. For example, it can differentiate between different types of brain tumors based on their unique imaging characteristics, such as signal intensity, contrast enhancement, and pattern of growth \citep{mazurowski2014computer}. This information is crucial for accurate diagnosis and guiding personalized treatment approaches, including surgery, radiation therapy, chemotherapy, and targeted therapies. 



\paragraph{Clinical prediction, radiogenomics, and multi-omics data integration} In the context of gliomas (brain cancer), MRI has been widely used to address several crucial questions of interest. One pivotal area of investigation involves the prediction of clinical outcomes, including survival, based on MRI-based tumor-derived features \citep{gutman2013mr,henker2017volumetric,zhou2017mri,bae2018radiomic,perez2018tumor,perez2019morphological}. Methods have also been developed to predict genomic profiles/molecular characteristics using radiological features extracted from MRI data \citep{darvishi2020prognostic,mohammed2022quantifying,liu2023radiogenomic}. 
Additionally, the integration of radiology data with genomic information, referred to as \textit{radiogenomics}, has emerged as a promising avenue for unraveling the complex molecular underpinnings of gliomas and identifying novel biomarkers associated with tumor behavior \citep{mazurowski2017radiogenomics,fathi2020imaging,singh2021radiomics}. Radiogenomic associations elucidate the relationship between imaging phenotypes and underlying genetic alterations, paving the way for precision medicine approaches that leverage both imaging and molecular data for prediction of clinical phenotypes \citep{nicolasjilwan2015addition}. Furthermore, there is growing interest in the integration of data across multiple platforms, such as MRI, genomics, proteomics, and clinical data, to comprehensively characterize the tumor microenvironment, gain deeper insights into the underlying molecular mechanisms, and identify interactions that drive disease progression \citep{antonelli2019integrating,boehm2022harnessing}. Such integration represents a powerful approach for advancing our understanding of cancer, with direct implications for personalized medicine and improved patient care. 

\begin{figure}[!t]
    \centering
    \includegraphics[width=\textwidth]{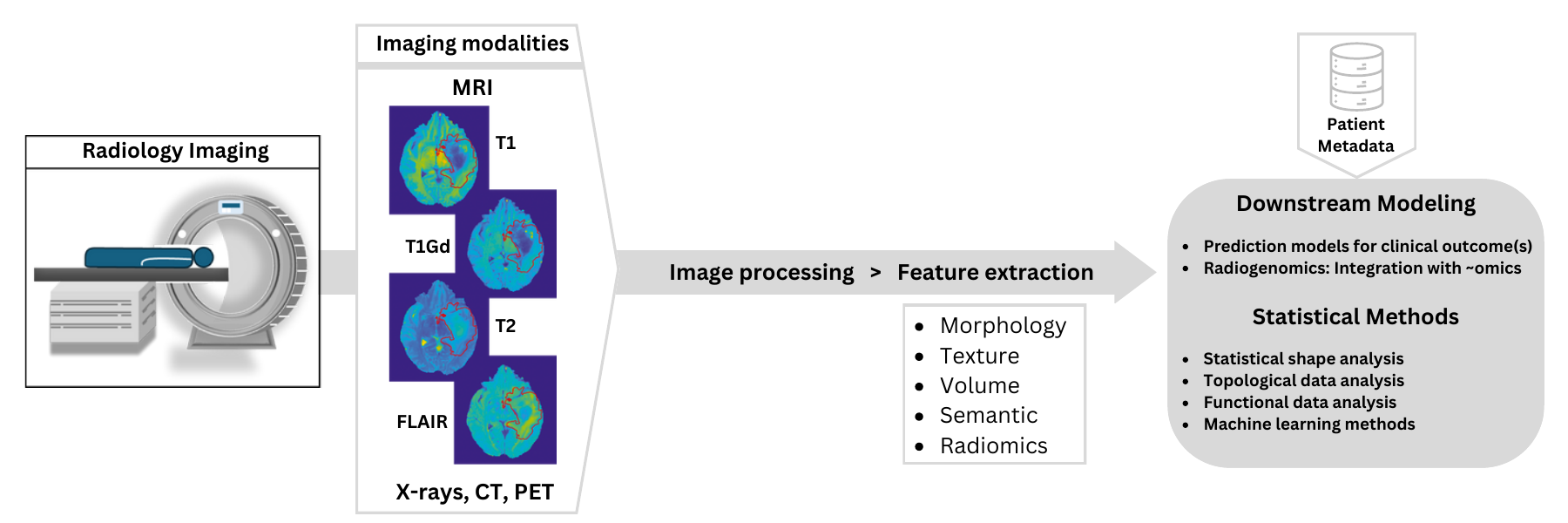}
    \caption{\textit{Workflow for radiology imaging data.} Radiology images, obtained as MRI, CT, or PET scans, undergo preprocessing for harmonization before feature extraction. These features primarily capture morphology and texture but vary depending on the application and characteristics of the region of interest targeted for capture. Downstream modeling utilizes these imaging features as covariates to predict clinical outcomes or as predictors/outcomes in models that integrate data from different platforms. Statistical methods predominantly focus on statistical shape analysis, functional and topological data analysis.}
    \label{fig:Radiology}
\end{figure}

In Figure \ref{fig:Radiology}, we present a comprehensive overview of the steps in a statistical modeling framework for radiology data. This includes image acquisition, data processing, feature extraction, and model building. In Section \ref{subsec:mri_feature}, we briefly discuss MRI data modalities and feature extraction techniques. Section \ref{subsec:mri_morphology} covers statistical methods for analyzing morphological features, employing tools from statistical shape analysis and topological data analysis. Section \ref{subsec:mri_texture} delves into the statistical methods used for texture features, with a focus on tools from functional data analysis. In Section \ref{subsec:mri_other}, we discuss statistical analysis methods specifically for volumetric features. Additionally, other related methods within these sections are presented as appropriate.

\subsection{MRI data and feature extraction}\label{subsec:mri_feature}

\paragraph{MRI modalities, radiomics and data processing}  MRI scans for gliomas employ various imaging \textit{modalities} (sometimes also referred to as \textit{sequences}) to capture detailed information about the tumor region. These sequences typically include T1, contrast-enhanced T1-weighted, T2, T2-weighted FLAIR (Fluid-Attenuated Inversion Recovery), and diffusion-weighted imaging (DWI). Each of these imaging sequences provide complementary information by highlighting/emphasizing regions with different characteristics (e.g., cerebrospinal fluid, peritumoral edema, necrosis) in the tumor. Clinicians can comprehensively characterize gliomas by integrating information from these different MRI sequences. \textit{Radiomics} involves the automated extraction and analysis of quantitative features from regions of interest in radiology images \citep{kumar2012radiomics,gillies2016}, with the aim of converting these images into multi-dimensional datasets for further analysis. In gliomas, radiomic features aim to capture a wide range of information, including texture, shape, intensity, and spatial relationships in the tumor region. The general process of any radiomic analysis consists of several sequential steps, including image acquisition and reconstruction, preprocessing, delineation of regions of interest, extraction and quantification of features, feature selection, and the development of predictive and prognostic models \citep{thawani2018radiomics}. To address challenges such as variability due to scanner differences between sites, various image processing procedures are employed, including intensity normalization, voxel intensity calibration, and bias field correction, as preliminary steps before radiomic feature extraction \citep{shinohara2014statistical,um2019impact,van2020harmonization}. The delineation of regions of interest (e.g., tumors) can be accomplished through manual, semi-automated, or fully automated methods \citep{pereira2016brain,bakas2017advancing}, followed by the extraction of radiomic features \citep{tiwari2016computer}. Note that as part of image data processing, some methods identify a specific two-dimensional slice of interest (e.g., the axial slice with the largest tumor area) and compute radiomic features based on this slice (e.g., the 2D shape of the tumor, considered as a curve). On the other hand, some methods utilize the entire 3D image or the 3D tumor region (e.g., the density of tumor pixel values from the entire tumor region). The choice between these options largely depends on whether the features used by a method can be constructed using 2D or 3D images. This review includes methods that use specific 2D slices as well as those that use the entire 3D images.



\paragraph{Radiomic feature extraction} In radiomic analysis, features can be broadly classified into distinct groups according to the aspects of the image or region of interest they aim to quantify. \textit{Morphological} features describe the shape and size characteristics of the tumor, e.g., circularity, lengths of major/minor axes, and elongation ratio, which provide insights into overall tumor shape. \textit{Texture} features capture spatial patterns and heterogeneity within the tumor region, e.g., histogram-based measures such as mean and skewness, homogeneity, and entropy \citep{just2014improving}. These features reflect variations in pixel intensity and distribution, offering information about tumor microenvironment and its composition. Additionally, grey level co-occurrence matrix (GLCM) features employ combinations of grey levels based on distance and angle values to characterize image texture \citep{haralick1973textural}. \textit{Volumetric} features focus on the three-dimensional characteristics of the tumor and its surroundings \citep{zhang2014identifying}, providing additional information about tumor morphology and spatial relationships with adjacent structures; examples include tumor and tumor sub-region volumes, and surface-to-volume ratio. Finally, qualitative \textit{semantic} features, such as tumor location, shape, and geometric attributes observed on structural MRI scans, are utilized by neuroradiologists to characterize the tumor environment \citep{zhou2018radiomics}. The Visually AcceSAble Rembrandt Images (VASARI) project, established by The Cancer Imaging Archive (TCIA), has defined a feature set to ensure consistent glioma description using specific visual features and controlled terminology \citep{rios2015fully}. Semantic features are robust to variations in image acquisition parameters and noise, and are usually integrated with more advanced radiomic features for modeling purposes. 

\paragraph{Machine learning for radiomics} In the realm of precision oncology, machine and deep learning algorithms are pivotal for mining vast medical imaging data to uncover complex biological patterns, aiding in personalized cancer diagnosis and treatment \citep{avanzo2020machine}. Radiomics modeling methods are bifurcated into {\it feature-engineered} (traditional radiomics) and {\it non-engineered} (deep learning-based radiomics), with further classifications as supervised, unsupervised, or semisupervised learning for both feature-based and featureless approaches. The process of feature selection, crucial to effective outcome modeling, involves identifying the most relevant subset of features that optimally correlates with endpoints. Various feature selection methods—embedded (e.g. least absolute shrinkage and selection operator, LASSO, \cite{tibshirani1996regression}), wrapper (e.g. support vector machines, SVM, \cite{suthaharan2016machine}), and filter methods—differ in their approach and interaction with machine learning algorithms. Classification (e.g., disease recurrence) and survival analysis (e.g., impact of treatment on survival time) are tackled with different machine learning models like logistic regression, SVM, and random forests. Deep learning, commonly used in a featureless context, is finding its way into feature-based applications, with neural networks enabling the exploitation of unlabeled data through semisupervised learning methods. This suite of tools, including advanced methods like Cox regression and random survival forests, provides a comprehensive framework for predictive modeling in cancer care \citep{ishwaran2008random,VANBELLE2011107}. However, in this review paper, we will instead focus on various advanced statistical modeling techniques that have been formulated to analyze mophological, texture and volumetric features.

\subsection{Morphology-based features}\label{subsec:mri_morphology}


Tumors typically exhibit variations in both physiological and shape-related attributes \citep{marusyk2012intra}, and understanding both intertumor and intratumor heterogeneity is crucial in characterizing them \citep{melo2013cancer}. The heterogeneity among tumors from different patients can be assessed through morphological characteristics such as size and shape, alongside genomic and clinical data \citep{mclendon2008comprehensive}. Several related methods rely on subjective features provided by experts and numerical summaries, which only partially capture tumor shape and suffer from reproducibility issues. In this section, we will focus on statistical methods for analyzing the shapes of regions of interest in radiology images, specifically focusing on elastic shape analysis and topological data analysis.

\subsubsection{Elastic shape analysis}\label{subsubsec:elastic}

Elastic shape analysis is a powerful tool to the study of geometrical properties and deformations of shapes \citep{srivastava2010shape}. Two comprehensive texts that present the theory and application of statistical shape analysis include \cite{dryden2016statistical} and \cite{srivastava2016functional}. In the following we present some examples of how this is applied in biomedical imaging data analysis to identify and quantify subtle morphological differences in anatomical structures, which can be indicative of clinical outcomes.

\cite{bharath2018radiologic} modeled tumor shapes as properties of parametric curves in $\mathbb{R}^2$, allowing flexibility in handling shape features. A shape space is defined where tumor shapes are represented as closed curves and distance between curves is defined using a geodesic path on a manifold. A shape-based principal component analysis is conducted to identify and visualize principal directions of variation in a sample of tumor shapes. \cite{bharath2018radiologic} demonstrates these tools in clustering glioblastoma tumor shapes on a cohort sourced from TCIA \citep{clark2013cancer} and The Cancer Genome Atlas (TCGA) \citep{mclendon2008comprehensive}. Additionally, they illustrate the utility of the developed tools in other inferential tasks such as two-sample testing and survival time modeling. This approach provides a coherent statistical representation of tumor shape by leveraging a geometric framework, and integrates tumor shape as a potential prognostic factor in statistical modeling. It can be applied beyond glioblastoma to other cancers and imaging modalities.

The shapes of regions of interest in biomedical images have generally been studied by elastic shape analysis. \cite{matuk2020biomedical} summarizes methods from geometric functional data analysis for modeling of shapes of curves and surfaces in various biomedical contexts. While considering curves and surfaces as objects, parameterization-invariant Riemannian metrics are developed using square-root transformations that simplify the geometry. This enables efficient computation of (i) Karcher mean of these complex data objects, and (ii) principal component analysis on appropriate representation spaces to explore the variability. Furthermore, elastic shape analysis \citep{srivastava2010shape,kurtek2012statistical,bharath2020analysis} provides tools that are invariant to translation, scaling, rotation, and reparameterization, to analyze shapes \citep{kendall1984shape}; an example will be described in Section \ref{subsec:mri_texture} to study the shapes of probability density functions constructed to assess the tumor texture. Note that while processing the images for elastic shape analysis, curve extraction often entails identifying the shape of regions of interest, such as tumor boundaries. This boundary is usually represented for computational purposes as a set of distinct two-dimensional points that outline the shape. However, this can introduce noise, particularly in low resolution scenarios, warranting careful considerations regarding the data inclusion.

There are other methods available for shape analysis including deformable models such as the Large Deformation Diffeomorphic Metric Mapping framework \citep{glaunes2008large}. This framework is used for diffeomorphic mapping, and it aims to assign metric distances on the space of images. It facilitates the direct comparison and quantification of morphometric alterations in shapes. This method is quite general and can accommodate different types of shape representations including point clouds, landmarks, curves, surfaces, and images, and has been used for the analysis of biomedical imaging data \citep{ceritoglu2013computational}.

\subsubsection{Topological data analysis}\label{subsubsec:topological}

In recent years, algebraic topology tools have gained popularity in shape analysis applications due to their ability to extract tractable algebraic invariants from complex shape data. In the context of images, topological data analysis (TDA) interprets them as point clouds, as opposed to considering them as pixel arrays. Central to TDA approach is persistent homology, which captures multiscale topological features of a shape by tracking the `births' and `deaths' of homological features as a filtration function increases in value \citep{carlsson2009topology,chazal2017high}. The resulting persistence diagram provides a summary statistic for the shape, facilitating its analysis in various applications \citep{carlsson2014topological,edelsbrunner2022computational}. Refer to \cite{carlsson2021topological} for a more comprehensive understanding of TDA.

\cite{crawford2020predicting} quantitatively analyzed glioblastoma MRI scans using TDA and assessed its clinical applicability. They introduced a novel statistic, the smooth Euler characteristic transform, which summarizes the tumor shapes as a collection of smooth curves. The key strategy is to construct these topological summary statistics as a function that maps shapes into a Hilbert space. This approach ensures one-to-one mapping and admits a well-defined inner product structure, facilitating the adaptation of functional data analysis concepts to incorporate shape summary statistics as predictors into regression models. Previous TDA methods such as the persistent homology transform produce similar summary statistics but are incompatible with functional data models due to a lack of inner product structure. This proves advantageous in predicting clinical outcomes, offering a robust approach for quantifying diverse tumor shapes and their impact on patient prognosis. \cite{crawford2020predicting} bridges topological analysis and regression modeling and enhances its utility in imaging research, particularly in elucidating complex relationships between tumor morphology and clinical outcomes. The software is available on GitHub\footnote{\url{https://github.com/lorinanthony/SECT}}. This framework was applied to TCIA/TCGA glioblastoma cohorts \citep{mclendon2008comprehensive,clark2013cancer}, and yielded insights into the relationship between shape features and survival outcomes influenced by molecular heterogeneity. 

A related work by \cite{jiang2020weighted} introduces the weighted Euler characteristic transform, a variation of the Euler characteristic transform tailored for shape data represented by a weighted simplicial complex. This transformation encapsulates both the shape and the weighting function in a topological summary. They construct a simplicial complex in $\mathbb{R}^2$ from the segmented shape and define a weighting function based on grayscale tumor pixel values from MRI scans. The software is available on GitHub\footnote{\url{https://github.com/trneedham/Weighted-Euler-Curve-Transform}}. Glioblastoma tumors are analyzed using these weighted Euler characteristic transform representations to cluster the tumors according to survival times.

\cite{wang2021statistical} extend the method in \cite{crawford2020predicting} to develop a pipeline that additionally enables variable selection with shapes as covariates. First, it summarizes the geometry of 3D shapes (represented as triangular meshes) by encoding changes in their topology into a collection of vectors or curves. Then, using nonlinear Gaussian process models, it classifies the shapes based on these topological summaries. It computes effect size analogs and association metrics for each topological feature used in the classification, providing evidence of their association with specific classes. Finally, through an iterative process, these topological features are mapped back onto the original shapes, highlighting the spatial locations that best explain the variation between groups. The software is available on GitHub\footnote{\url{https://github.com/lcrawlab/SINATRA}}. This framework was applied to analyze mandibular molar shapes from CT scans of different primate suborders, showcasing its ability to recover known morphometric variations across phylogenies.

Another related work by \cite{somasundaram2021persistent} explores the application of persistent homology via TDA to characterize lung cancer and predict survival outcomes. They used cubical complexes to quantify topological features at varying pixel intensities. Cubical complexes are a series of binary images generated to effectively analyze the number of topological features in an image as the image is thresholded at different intensities. This generates a novel output which is termed \textit{the feature curve}, and its first moment serves as a summary statistic. This summary was associated with survival, adjusting for tumor size, age, and stage, when applied to segmented CT lung scans from two datasets. The software is available on GitHub\footnote{\url{https://github.com/eashwarsoma/TDA-Lung-Phom-Reproducible}}. Results indicate that lower first moment scores are significantly linked to better survival outcomes, suggesting the potential utility of persistent homology features in clinical oncology decision-making alongside traditional radiomics variables.

\cite{moon2023using} introduced a topological feature computed via persistent homology to characterize tumor progression from MRI scans and assess its impact on time-to-event data. A distance transform is applied on tumor shapes, facilitating the computation of persistent homology. This method quantifies all tumor shape patterns, irrespective of size, and remains invariant to transformations like rotation and translation. The topological features are represented in a functional space and used as functional predictors. Using functional Cox proportional hazards modeling, incorporating clinical variables and functional persistent homology features, the study provides interpretable fundings on the association between tumor shape patterns and survival outcomes \citep{yao2005functional,chen2015statistical}. The software is available on GitHub\footnote{\url{https://github.com/chulmoon/TopologicalTumorShape}}. Results from a case study on brain tumor patients indicate that these topological features predict survival prognosis, with high-risk groups exhibiting worse outcomes, particularly characterized by irregular and heterogeneous shape patterns indicative of aggressive tumor progression. Overall, the method enables detailed shape and pattern analysis in predicting survival using topological features extracted from MRI scans.

\cite{oyama2019hepatic} applied similar ideas to assess the accuracy of hepatic tumor classification using MRI scans through TDA and machine learning prediction models. They evaluated hepatic tumors from three subtypes and created persistence diagrams (of degree 0, degree 1, and degree 2), providing a multiscale description of topological features. This study underscores the potential clinical utility of TDA in computer-aided diagnosis of tumors, offering a non-invasive approach for diagnosis using radiology imaging.

\subsection{Texture-based features}\label{subsec:mri_texture}

Biomedical images often display heterogeneity in regions of interest, arising from the physiological and morphological properties of the tissues under examination \citep{marusyk2012intra}. In radiology images, texture features describe various aspects of pixel intensity variations and are crucial in assessing the heterogeneity (or homogeneity), structure, composition, and organization of tissues (or tumors). Numerous machine learning techniques are based on the construction of prediction models by extracting summary features from pixel intensity values within regions of interest \citep{singh2021radiomics}. However, in this section, our primary emphasis is on statistical frameworks designed to rigorously quantify heterogeneity, which are then integrated into downstream modeling for prediction and inference purposes. Several approaches involve constructing features that can be considered as functional data objects, leading to the development of methods capable of using them as either predictors or outcomes. We will first outline methodologies for analyzing texture features within the functional data framework. Then, we will introduce studies using other advanced statistical methods.

\subsubsection{Functional data analysis}\label{subsubsec:functional}

Machine learning approaches have primarily focused on analyzing physiological attributes at the pixel or voxel level by first summarizing these attributes from regions of interest into `summary statistics' or metrics \citep{singh2021radiomics}. For example, metrics such as skewness, kurtosis, and entropy of voxel intensities are used, but these metrics are often selected subjectively, resulting in loss of information about heterogeneity. Recent works have thus focused on harnessing the entire distribution of the intensities for statistical modeling. A number of these approaches employ techniques from functional data analysis. For further details on functional data analysis, reference books include\cite{ramsay2005functional} and \cite{srivastava2016functional}.

\paragraph{Clustering distributions} \cite{SAHA2016132} introduced a novel approach that generates density profiles of voxel intensities for each patient's segmented tumor region to comprehensively summarize the tumor heterogeneity. These densities are used directly as data objects. A Fisher-Rao metric is formulated to measure the similarity (or dissimilarity) between density profiles, allowing comparisons of tumor heterogeneity between different patients. This distance metric is then used to create tools for computing the mean of a set of density functions and conducting geometric principal component analysis (PCA) for dimension reduction within a Riemannian geometric framework. This framework enables comparisons and clustering of patients based on their density profiles. The key innovation lies in using the entire distribution of tumor intensities, rather than summarizing them into histograms or summaries, as a representation of tumor heterogeneity. Application of this methodology to MRI images from the TCGA glioblastoma dataset revealed significant clusters of patients corresponding to different anatomical characteristics of the tumor, indicating varying levels of disease aggressiveness. These clusters were validated against known molecular subtypes, genomic signatures, and prognostic clinical outcomes using imaging biomarkers, yielding both new insights and confirming previous findings.


\paragraph{Distributional regression}  \cite{yang2019quantile} developed a method called quantile function regression, which models the complete marginal distribution of pixel intensities as functional data, expressed through the quantile function. These subject-specific quantile functions were regressed against demographic, clinical, and genetic predictors. This enables assessment of the global association of covariates with the distribution and identify distributional characteristics such as mean, variance, skewness, heavy-tailedness, and various upper and lower quantiles that characterize these differences. A custom basis function set known as `quantlets' was introduced to address smoothness in the functions, intrafunctional correlation, and to enhance statistical power. Quantlets are sparse, regularized, nearly lossless, and empirically defined, adapting to the features of the dataset. The model is fit under a Bayesian framework incorporating nonlinear shrinkage of quantlet coefficients to regularize functional regression coefficients and provide fully Bayesian inference using Markov chain Monte Carlo after fitting. The method was applied to analyze tumors from MRI scans from the TCGA glioblastoma dataset. The findings revealed specific differences in tumor pixel intensity distribution between genders and tumors with and without DDIT3 mutations, highlighting the utility of quantile functions in capturing tumor heterogeneity.


Advanced statistical frameworks to assess tumor heterogeneity have facilitated exploration of the relationship between molecular composition and tumor heterogeneity by integrating radiology imaging with genomic data, known as \textit{radiogenomics}. \cite{mohammed2021radiohead} developed a distributional regression framework (called {\small RADIOHEAD}) to uncover radiogenomic associations by integrating MRI and genomic data. Tumor heterogeneity is represented as tumor voxel-intensity probability density functions. For each patient, density functions are generated for three subregions of the tumor: necrosis and non-enhancing tumor, edema, and the enhancing core. A Riemannian-geometric framework on the space of probability density functions is utilized to conduct a principal component analysis to map these density functions of tumor subregions to principal component scores. These scores are then used as predictors in a Bayesian regression model, with pathway enrichment scores (obtained by gene set variation analysis) derived from gene expression profiles as the response variable. Variable selection is conducted using a group spike-and-slab prior by leveraging the group structure among predictors induced by tumor subregions. Bayesian false discovery rate analysis is subsequently used to identify significant associations based on the posterior distribution of regression coefficients. The software is available on GitHub\footnote{\url{https://github.com/shariq-mohammed/RADIOHEAD}}. The methodology is applied to the lower grade glioma (LGG) dataset from the TCGA/TCIA cohort. The findings highlight several pathways relevant to LGG etiology as having significant associations with imaging-based predictors.

\cite{mohammed2023tumor} developed another statistical framework for uncovering radiogenomic associations within LGGs. This approach involves creating a unique imaging phenotype by partitioning the tumor region into concentric spherical layers, mimicking the tumor evolution process and enabling a nuanced analysis of genomic associations across different tumor regions. Within each layer, MRI data is represented by voxel-intensity-based probability density functions, offering a comprehensive view of tumor heterogeneity. Employing a Riemannian-geometric framework, these densities are transformed into principal component scores, generating layer-specific imaging phenotypes. The overall model framework is a multivariate-multiple regression setup that integrates structural attributes from imaging and genomic data, where imaging phenotypes are treated as responses and genomic markers as predictors. Bayesian variable selection is conducted leveraging a continuous structured spike-and-slab prior for each layer. The hierarchical prior formulation accommodates the interior-to-exterior structure of layers and the correlation between genomic markers. A scalable Expectation-Maximization-based estimation strategy is developed to efficiently estimate model parameters, facilitating model selection and scalability. The software is available on GitHub\footnote{\url{https://github.com/shariq-mohammed/marbles}}. This approach enables robust identification of radiogenomic associations, contributing to the understanding of LGG etiology and potential diagnostic markers. Although it was developed for imaging-genomics, the framework offers a general sequential modeling approach applicable to various domains with natural ordering.


\paragraph{Other applications of functional data analysis} Gray-level co-occurrence matrices (GLCMs) have been used to study spatial dependencies among gray levels within delineated regions of interest in radiology images \citep{narang2017tumor}. A GLCM is a matrix computed over an image domain, with its entries representing counts of co-occurring gray-scale values in image voxels at specified spatial offsets and orientations. While many analyses focus on summary statistics derived from GLCM (such as contrast, correlation, energy, entropy, and homogeneity), \cite{chekouo2020bayesian} proposed a Bayesian framework to model associations between a clinical outcome and GLCMs constructed from tumor regions in MRI scans. This is accomplished by treating the GLCM as a two-dimensional functional data object rather than using traditional analysis methods that rely on summary statistics computed from GLCMs as texture features, enabling a deeper understanding of spatial dependencies in gray levels. To accommodate the spatial structure inherent in GLCMs, a generalized two-dimensional functional linear model is employed within a Bayesian framework. The software is available on GitHub\footnote{\url{https://github.com/chekouo/BayesF2D}}. This approach provides a clear interpretation and enables effective modeling of scalar outcomes with functional predictors to analyze complex data relationships. It offers a novel perspective on tumor heterogeneity estimation, leveraging the spatial information in GLCMs to enhance predictive performance. It was implemented on the TCIA LGG dataset to predict isocitrate dehydrogenase (IDH) mutation status.

The T2-FLAIR mismatch sign \citep{patel2017t2} in LGGs is a reliable diagnostic tool to identify a specific molecular subtype, namely IDH-mutant 1p/19q noncodeleted astrocytomas. IDH mutations are associated with improved survival, while 1p/19q codeletion indicates better survival outcomes and increased sensitivity to treatment. The T2-FLAIR mismatch sign is characterized by complete/near-complete hyperintense T2 signal and a relatively hypointense FLAIR signal, except for a hyperintense peripheral rim. \cite{mohammed2022quantifying} developed a statistical framework to quantify this sign and predict the molecular status of LGG tumors from TCIA. This approach employs geographically weighted regression within segmented tumor regions to objectively estimate the T2-FLAIR mismatch. A probability density function was constructed using residuals from the geographically weighted regression, allowing subsequent statistical analysis. A permutation-based hypothesis test was developed to test for differences in residual signatures between different groups of interest (based on clinical outcomes) and to develop classification models to predict different molecular subtypes of LGGs, including IDH-mutant 1p/19q noncodeleted astrocytomas, based solely on non-invasive MRI scan markers. 

Integrative radiogenomic analysis, which combines radiology imaging with genomic data, provides valuable insights into disease mechanisms. However, accurately estimating uncertainty in these integrative models requires addressing the bias introduced by variable selection. \cite{panigrahi2023integrative} proposed a selection-aware Bayesian framework to mitigate this bias and improve inference accuracy. This approach incorporates strategies from selective inference, integrating a selection-aware posterior into flexible integrative Bayesian models to correct for the impact of variable selection. The methodology involves a two-stage integrative modeling framework, progressing from genomics to imaging to clinical outcomes. In the first stage, a candidate set of genomic variables associated with intermediary imaging phenotypes is identified. These imaging phenotypes are constructed as probability density functions of tumor pixel intensities to capture tumor heterogeneity. In the second stage, variables are selected from an imaging-informed set of genomic variables associated with clinical outcomes. This model setup mirrors the natural progression of cancer from genomic changes and physiological characteristics to clinical manifestations.

\cite{cho2022tangent} investigated the relationships between nonlinear functional data extracted from the same object, that is, the shape of tumors and the density function of tumor pixel intensities from MRI scans. They propose a method for canonical correlation analysis between these paired probability densities and shapes of closed planar curves. This approach adopts a Riemannian geometric setting for the non-Euclidean spaces of probability density functions, using the Fisher–Rao metric under a square root transformation. Elastic shape analysis is used for the shape curves, measuring changes via an elastic metric using a square root velocity function. The method combines linearization and dimension reduction of the data using tangent space coordinates, using the tangent space parameterization of the resulting manifolds at a fixed point. Dimension reduction is achieved by projecting locally linearized functional data objects onto a finite-dimensional subspace spanned by eigenbasis functions. The resulting coefficient vectors serve as finite-dimensional Euclidean representations of density functions and shape curves, enabling standard multivariate canonical correlation analysis. Canonical variate directions are computed and visualized directly on the space of densities and shapes. The method is applied to a glioblastoma dataset from the TCIA cohort to analyze the interactions between tumor texture and morphology.

\subsubsection{Tree-based and spatial methods}\label{subsubsec:othertexture}

\cite{bharath2017statistical} investigated tumor heterogeneity in TCGA glioblastoma patients to discern variations in heterogeneity between those with long and short survival times. They create a novel quantifiable representation of tumor heterogeneity and cluster participants based on survival times. Unlike traditional methods summarizing tumor pixel intensities, they proposed constructing binary trees from the tumor regions in MRI scans. Employing an agglomerative hierarchical clustering algorithm on tumor pixel intensities, the total path length for randomly chosen subsets of leaves was computed. These lengths served to quantify the distances between clusters of similarly valued intensities, effectively capturing the heterogeneity within the tumor image. A novel hypothesis test tailored for trees was developed to analyze differences between glioblastoma patients with long and short survival times. The findings revealed distinct clusters, with images from patients with shorter survival times showing more varied clustering, indicative of greater tumor heterogeneity. 

\cite{eloyan2020tumor} presented a statistical framework for estimating tumor heterogeneity by incorporating the spatial context of voxel intensities. The approach identifies clusters of voxels with similar intensity profiles using Markov random fields to model voxel intensities, thereby accounting for their spatial distribution. The within and between cluster summary measures of the clusters serve as descriptors of tumor heterogeneity, focusing on cluster sizes, intensities, and a shapes. Applying this method to predict survival times for lung cancer patients using TCIA CT scans, it outperforms established methods like Cox proportional hazards and generalized linear models with various penalization techniques. The findings showcase a novel approach that accounts for spatial voxel distribution and demonstrates superior predictive performance in lung cancer survival prediction.

\cite{li2019spatial} developed a Bayesian probabilistic approach for predictive classification of GLCM objects using Gaussian Markov random fields. This method incorporates spatial dependencies among GLCM entries using a spatial Gaussian process, establishing distributional assumptions for the normalized GLCM space. The Bayesian spatial Gaussian process classifier demonstrated efficacy when applied to distinguish malignant and benign adrenal lesions in CT scans.
In \cite{li2021bayesian}, a Bayesian nonparametric framework is introduced for inferring the GLCM as a multivariate count object, capturing inherent spatial dependence patterns within GLCM lattices. Model specification is achieved through a rounded kernel mixture model, assuming the spatial distribution of GLCM cell counts originates from a latent continuous random vector defined on the GLCM lattice. This framework enables flexible and robust estimation of the data generating distribution for count data, considering both intralesion heterogeneity of GLCM objects and spatial correlation of lattice counts. The multivariate latent vectors are modeled using a spatial Dirichlet process with a conditionally autoregressive Gaussian structure, effectively addressing over and under dispersion of observed counts while accounting for spatial autocorrelation among nearby cells in the GLCM lattice.

\subsection{Volumetric features}
\label{subsec:mri_other}
Radiomic analyses have primarily focused on shape and texture-based features. However, the tumor volume and its constituent parts (such as necrosis, enhancing, and nonenhancing regions) are critical factors considered during clinical assessment, with studies highlighting their prognostic significance for the overall survival of glioblastoma patients \citep{iliadis2012volumetric}. It is of interest to understand the combined influence of genes and identifying pathways (groups of genes) that demonstrate notable associations with volume-based imaging characteristics. \cite{chekouo2023bayesian} developed a method to elucidate genomic markers collectively impacting the composition of tumor components. The study examined compositional outcomes, specifically the volumes of tumor subregions, including necrotic, edema/invasion, contrast-enhancing, and nonenhancing regions. A Bayesian hierarchical model was developed for variable selection within a group framework, specifically tailored for correlated multivariate compositional response variables. This model utilized a Dirichlet distribution to represent the tumor composition data, facilitating the integration of available high-dimensional covariate information within a log-linear regression framework. For each outcome (subregion volume), a spike-and-slab prior was employed to select variables at both the group level and within each group. Notably, this prior differed from other spike-and-slab priors for group selection by incorporating a one-layer indicator variable for group selection and accounting for the overlapping structure between groups, reflecting the biological reality where multiple genes may be involved in different pathways. The software is available on GitHub\footnote{\url{https://github.com/chekouo/oBGSelComp}}. The approach was applied to the TCGA glioblastoma cohort, revealing several genomic factors contributing to the observed heterogeneity in the composition of brain tumors.



The literature surveyed predominantly covers statistical modeling methodologies designed for analyzing radiomic features within cancer research, primarily leveraging MRI or CT scans. These studies have primarily focused on developing methods for analyzing features associated with morphology, texture, and volumetrics, whether for prediction modeling or data integration purposes. However, this focus does not fully capture the breadth of statistical approaches relevant to radiology imaging datasets. Different diseases may require distinct and unique imaging features, underscoring the importance of exploring alternative statistical methodologies tailored to specific diagnostic contexts within the broader realm of biomedical imaging.

\section{Pathology Imaging}\label{sec:Pathology_imaging}

\subsection{Quantitative digital pathology}

Digital pathology is the study of digitized, high-definition images of tissue samples. In recent years, such images have become the object of a great deal of quantitative and computational research \citep{Baxi22}. This research generally pursues one of two overarching goals: automation of existing pathological assessment (e.g., cell classification and assessment of tumor stage) or assessment of the tumor and tumor microenvironment (TME). While the former goal tends to intersect broadly with machine learning, the latter is an active area of both machine learning and statistical research. Accumulating evidence suggests that the spatial composition of the TME has implications for both patient prognosis and potential treatments \citep{Sadeghi21, Falcone20}. This fact, along with the granularity of spatial analysis enabled by modern digital pathology, has sparked substantial interest in the modeling and analysis of the TME. Similarly, interest in assessing prognostic information from the shapes of tumors has sparked research into statistical shape analysis as applied to imaging data \citep{Zhang24}.

Quantitative digital pathology is extremely broad, and intersects deeply with medicine, computer science, and biostatistics. Given this, a complete review would be beyond the scope of this paper. For the purposes of this review, we limit the scope to methods we consider to those that are statistical in nature, in either their primary formulation or their downstream modeling. We consider these methods in the application context of \textit{Hematoxylin and Eosin} (H\&E) stained images.

Figure \ref{fig:HandE} illustrates an overview of the typical analysis workflow with H\&E images. After the images are prepared (Section \ref{sec:HandEimageprep}), they are generally pre-processed in one or more of three ways: semantic segmentation, patch segmentation, or cell segmentation (Section \ref{sec:HandEsegmentation}). The resulting data is then modeled and analyzed depending on the type of preprocessing that was done (Sections \ref{sec:HandEsummarybased} and \ref{sec:HandEmodelbased}). The results of this modeling are generally used in downstream analysis.

\subsection{H\&E image preparation}\label{sec:HandEimageprep}

\begin{figure}[!t]
    \centering
    \includegraphics[width=\textwidth]{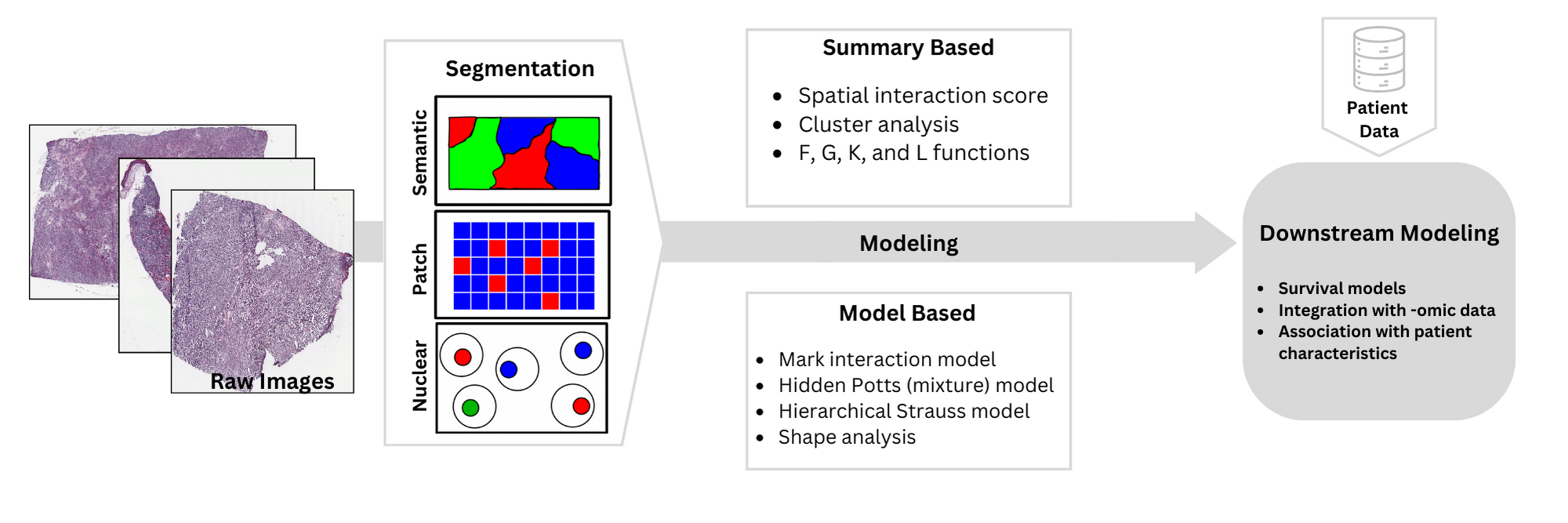}
    \caption{\textit{Workflow for H\&E imaging data.} Raw high-resolution images of tissue samples are segmented, with the method of segmentation depending on the analysis that will be performed. The segmented data is then analyzed via a summary based method or a model based method. The results of this model are then used in various forms of downstream analysis.}
    \label{fig:HandE}
\end{figure}

The starting point for any histopathological analysis is tissue samples. Sections of tissue are resected from regions of interest, at which point the preparation of the tissue for pathological assessment begins. At a high level, this process involves the \textit{fixation} of tissue to prevent post-resection corruption of the sample, \textit{embedding} of fixed tissue into paraffin blocks, \textit{sectioning} of the tissue (i.e. slicing into thin tissue sections), and \textit{staining} of the resulting slices \citep{Slaoui17}. For this review, we will limit our consideration to H\&E stained images, which is the de facto standard staining method for histopathological samples. H\&E staining colors the nuclei of the cells in a biopsy blue, and the cytoplasm and extracellular matrix pink \citep{Chan14}. This combination results in the distinctive purple appearance of H\&E stained biopsies, as seen in the raw images in Figure \ref{fig:HandE}. This staining is key to the analysis of histopathological images, as it more readily enables identification and analysis of the nuclei of cells.

Once the sections of the biopsy have been stained, they are then digitized using highly specialized scanners. While there are many methods and tools used to achieve this goal, the defining feature is the usage of an optical microscope to digitize the entirety of a stained tissue section \citep{Li22}. The result is an extremely high-definition image of an H\&E stained pathological tissue sample, which serves as the foundation for all subsequent quantification, modeling, and analysis. 


\subsection{Preprocessing and segmentation} \label{sec:HandEsegmentation}

Once the tissue sections have been processed into images, they require additional preprocessing before they are suitable for statistical analysis. While the exact nature of this preprocessing will vary depending on the context, it generally, at a minimum, entails some degree of \textit{image segmentation}. Image segmentation is broadly concerned with partitioning the pixels of an image into discrete classes \citep{Li22}. In the context of whole slide imaging, the goals of image segmentation range considerably, but generally fall into three categories: `patch' segmentation, semantic segmentation, and nuclear segmentation. Patch segmentation is when discrete sub-regions of an image of pre-specified size are classified, e.g., \cite{Kwok18} trained a convolutional neural network to classify patches of breast tissue as normal, benign, in-situ carcinoma, or invasive carcinoma. In semantic segmentation, on the other hand, a model is tasked with dividing up a whole slide image into some number of pre-defined classes. For example, \cite{Mehta18} trained such a model to segment whole slide images of breast cancer into seven different regions, including benign epithelium, normal stroma, and necrotic tissue. Finally, nuclear segmentation refers to models that delineate nuclei of cells from the surrounding cytoplasm and, if necessary, from each other as in the ConvPath pipeline \citep{Wang19}.

Of these three types of segmentation (second step of Figure \ref{fig:HandE}), the ones that most commonly precede statistical analysis of H\&E imaging data are patch and nuclear segmentations. In analyses where nuclear segmentation is involved, resulting segmented data are commonly used to determine the nucleus centroid location (used as a proxy for cell location) and to identify the cell type. For example, in addition to nuclear segmentation, the ConvPath pipeline determines the locations of the centroids of the nuclei and classifies each cell as either a tumor cell, a stromal cell, or a lymphocyte. 

\subsection{Feature extraction and summary-based methods} \label{sec:HandEsummarybased}

Many analysis methods rely on quantifying and extracting a specific feature of a given biopsy without positing an explicit model for the underlying spatial data. A common target for this analysis is the \textit{interaction} between tumor cells and other cells in the TME. Interaction (also referred to as \textit{co-localization}) between tumor cells and lymphocytes is considered an important prognostic indicator in cancer diagnoses \citep{Tsujikawa20}. Hence, much of the analysis of H\&E images seeks to identify and quantify the degree of interaction in a given biopsy. This quantification happens primarily via summary-based methods, as detailed in this section, and model based methods, as detailed in Section \ref{sec:HandEmodelbased}.

\cite{Feng23} sought to quantify the degree of interaction between tumor cells and lymphocytes in lung adenocarcinoma. Whole slide images of biopsies were divided into non-overlapping square sub-regions using a sliding window method. The nuclei of each cell in each sub-region were then segmented, and the type (tumor cell, lymphocyte, stromal cell, or other) and locations were determined. In each sub-region, a graph was constructed by treating each tumor cell and lymphocyte as a node, and considering two nodes to have an edge between them if they were within 60 pixels (approximately 15 $\mu$m) of one another. The Tumor-Lymphocyte Spatial Interaction score for a given sub-region (TLSI$_{patch}$) was then computed as the ratio of the number of edges between tumor cells and lymphocytes to the number of tumor cells. Sub-regions were subsequently classified as either TLSI$_{patch}$-low or TLSI$_{patch}$-high. Finally, TLSI scores were computed at the biopsy level, whereby biopsies were considered TLSI-high if at least 85\% of the sub-regions were classified as TLSI$_{patch}$-high, and TLSI-low otherwise.

\cite{Saltz18} employed a somewhat similar method to a multitude of cancer sub-types, with a few key differences. First, they segment images according to necrosis to improve the ability to identify lymphocyte infiltration. Second, rather than classifying individual cells, they focus on classifying 50$\mu\text{m}\times 50\mu$m patches as either having lymphocyte infiltration or not. To quantify interaction at the biopsy level, they partitioned biopsies into 50$\mu\text{m}\times 50\mu$m patches, and used their model to determine which patches have lymphocyte infiltration. They then summarize the results for each biopsy in terms of the overall proportion of patches with lymphocyte infiltration and various clustering measures that capture the spatial nature of the lymphocyte infiltration. 

\cite{Chang16} took a different approach. In addition to using non-convolutional neural network-based methods for segmentation (specifically, k-means and landmark-based spectral clustering), they focus on modeling cells in sub-regions as spatial point processes. They apply several traditional methods of clustering analysis from spatial statistics: the F, G, K, and L functions. The F and G functions quantify the likelihood of observing a cell within a specified radius of another cell, while the K and L functions quantify the relative intensity of points at various radii compared to what would be expected by chance alone. These functions and their multi-type variants are commonly applied in the spatial multiplex imaging literature; see Section \ref{sec:spatialmultiplex} for further details.

\subsection{Model-based methods} \label{sec:HandEmodelbased}

Recent work has quantified the interaction between different cell types by specifying an explicit statistical model, predominantly spatial, for the underlying data. For example, \cite{Li19a} treat segmented and classified cells as a marked point process and model it using a Bayesian mark interaction model. This model conditions on the cell locations in a biopsy and defines a graph on the cells, where each cell is a node, and cells within a certain distance are considered to have an edge between them. Conditional on this structure, the model treats the types of cells as random. They specify an `energy function' in which the abundances of each cell type as well as their tendency to be adjacent to cells of other types are explicitly parameterized. This energy function is used to define a full data likelihood, which yields a distribution over all possible point types for each point. The estimated interaction parameters can then be used in downstream modeling as measures of interaction between the different types of points. The software is available on GitHub \footnote{\url{https://github.com/liqiwei2000/BayesMarkInteractionModel}}.

\cite{Li19b} developed and utilized the hidden Potts mixture model (HPMM) to quantify the interaction between different cell types. This followed \cite{Li17}, who had previously applied a hidden Potts model (HPM) to cancer imaging data. However, as \cite{Li19b} noted, the standard hidden Potts model assumes homogeneity of interaction across the entire biopsy, which is likely not satisfied in practice due to the heterogeneity of the TME. Whereas the mark interaction model relies on defining a graph directly on the cells in an image, both the HPM and HPMM define an auxiliary lattice, the vertices of which are modeled using either the Potts model or the mixture Potts model, respectively. The key difference is that the HPMM defines $k$ sets of interaction parameters for $k$ distinct sub-regions of the data which capture the tendency of cells of different types to interact in each of those distinct sub-regions. On the other hand, the traditional hidden Potts model only defines one set of interaction parameters. Note that while the number of sub-regions is set a priori, the classification of different areas of the underlying image is done during model fitting. The software is available on GitHub\footnote{\url{https://github.com/liqiwei2000/BayesHiddenPottsMixture}}.

\cite{Osher23} also seeks to quantify interaction at the biopsy level while accounting for within tissue-heterogeneity. To this end, each biopsy image is partitioned into non-overlapping sub-regions, each of which is modeled as a Hierarchical Strauss model (HSM), the density of which is given by $f(\bm{x}_1, \bm{x}_2) \propto 
\exp\left(n_1 \beta_1  + n_2 \beta_2 +
S_{R_{12}}(\bm{x}_1, \bm{x}_2) \gamma_{12}\right)$, where $\bm{x}_1$ is the set of $n_1$ observed tumor cells and $\bm{x}_2$ is the set of $n_2$ observed immune cells. $\beta_1$ and $\beta_2$ capture the first-order intensities of tumor cells and immune cells, respectively. The key parameter of interest is $\gamma_{12} \in (-\infty, \infty)$, which captures the degree of interaction between tumor cells and immune cells in a given sub-region. The HSM differs from the previously discussed models in that it treats both the types and locations of the cells being modeled as random. Additionally, it models the locations of the immune cells conditionally on those of the tumor cells; this conditioning is what distinguishes the Hierarchical Strauss model from the standard Strauss model. These model fits are used to characterize a novel measure of interaction, Cell Type Interaction Probability (\textit{CTIP}). Because a model is fit on each of the sub-regions of a given biopsy, the estimates of CTIP across sub-regions are summarized at the biopsy level for use in downstream modeling. Among other findings, biopsy level mean CTIP was found to be significantly associated with hazard of death. The SPARTIN software as well as a visualization tool is available on GitHub\footnote{\url{https://github.com/bayesrx/SPARTIN}}.

Unlike much of the work discussed thus far, \cite{Zhang24} used H\&E imaging data to investigate the boundaries of tumors and the association with patient prognosis. To this end, they introduce the Bayesian Landmark-based Shape Analysis (BayesLASA) model. BayesLASA models the segmented boundaries of H\&E images as closed polygonal chains, i.e., a list of points where the first is equal to the last. From this list of points, some subsets are selected as specific points where the path changes substantially. These points are referred to as landmarks, and are used to capture the heterogeneity in the boundary of the tumor. Measurements of this heterogeneity were then used in downstream modeling and assessed for association with patient survival. Among other things, the number of landmarks selected and the skewness of the distribution were both significantly associated with hazard of death. The software is available on GitHub\footnote{\url{https://github.com/bougetsu/BayesLASA}}.

We conclude this section by noting that many of the methods discussed throughout Section \ref{sec:spatialmultiplex} can also be applied to H\&E stained images, as in \cite{Chang16}. However, because they are more commonly applied to multiplex imaging data, we have left their explanation to Section \ref{sec:spatialmultiplex}.

\section{Spatial Multiplex Imaging}\label{sec:spatialmultiplex}


\paragraph{Imaging technologies} Multiplexed tissue imaging (MI) is a new and rapidly evolving technique that enables researchers to analyze both the prevalence of cell types and the spatial relationships between them in a tissue biopsy. The term multiplexing indicates the ability to detect multiple biomarkers simultaneously on a single tissue section. Multiplexed imaging is a broad category of technologies. High-resolution images generated from MI technologies are relatively cheap to acquire, enabling the generation of large datasets to study the associations of cellular spatial relationships with tumor growth, metastasis, drug resistance, and patient survival \citep{steinhart_spatial_2021,schurch_coordinated_2020}.


PhenoImager, formerly known as Vectra, is a popular technology for MI data generation. It belongs to the class of MI technologies called multiplex immunofluorescence (mIF). The data acquisition pipeline begins with the placement of a tissue biopsy on a slide for imaging. Multiple antibodies or markers are applied to stain the slide. The markers are labeled with fluorophores which emit light \citep{cancers13123031}. The multiplexed slides are then imaged using multispectral imaging.   

There are various other technologies which generate similar data; multiple biomarkers measured at the single-cell level with retained spatial information. These include PhenoCycler, formerly known as CO-Detection by Indexing (CODEX, \cite{kuswanto2023highly}), Multiplexed ion beam imaging by time-of-flight (MIBI, \cite{angelo2014multiplexed}), Imaging Mass Cytometry (IMC, \cite{giesen2014highly}), Matrix-assisted laser desorption ionization mass spectrometry imaging (MALDI-MSI, \cite{aichler2015maldi}), and Digital spatial analysis (GeoMx/DSP/CosMx, \cite{merritt2020multiplex}). \citet{van_dam_multiplex_2022} provides a detailed overview of these technologies.  Due to the popularity of mIF data, we will focus on the statistical considerations of this data type with respect to preprocessing and phenotyping in Section \ref{subsec:preproc}. The methods outlined in Sections \ref{subsec:first},\ref{subsec:second} are generally applicable to any MI technology.  

\begin{figure}
    \centering
    \includegraphics[width=16cm]{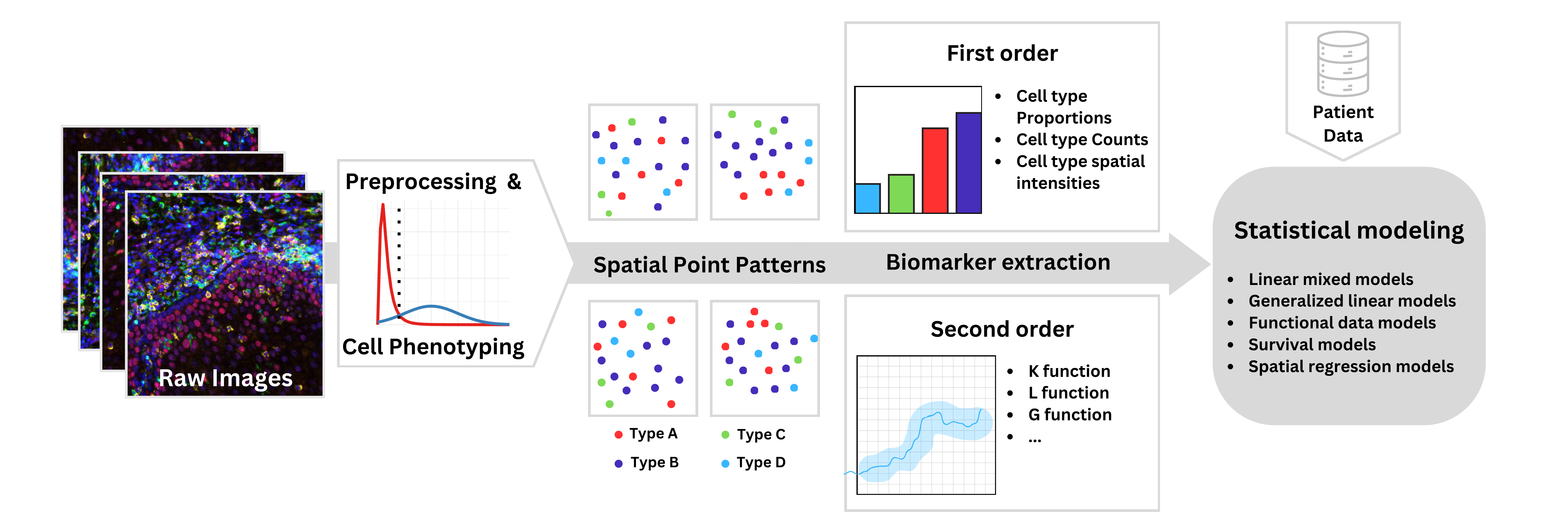}
    \caption{\textit{Workflow for MI data.} Raw high-resolution images of cells are segmented. The intensity of phenotypic markers are used to define the cell types (Type A, Type B, Type C, Type D, for example). First- and second-order biomarkers are extracted for each image and can be used in statistical modeling and analysis. (Credit for raw image: Allen Lab, NCI/NIH)}
    \label{fig:MI}
\end{figure}

\subsection{Preprocessing and phenotyping}\label{subsec:preproc}

Once the images are obtained, a series of preprocessing steps are applied to the raw images. The PhenoImager system uses proprietary software called InForm to process the images and phenotype the cells. An open-source and platform-agnostic alternative to InForm is Qupath \citep{bankhead2017qupath}. First, the single cells must be segmented. After segmentation, various processing steps are applied to facilitate cell phenotyping. The statistical considerations involved in the preprocessing and phenotyping steps are explored in detail by \citep{Wrobel2023}. To summarize, before the cells can be phenotyped, the raw images should be transformed, normalized, and batch corrected. Image transformations make the data more amenable to modeling, image normalization makes the marker distributions more similar across images, and batch correction removes non-biological sources of error \citep{harris2022quantifying}. Figure \ref{fig:MI} provides a graphical overview of the analytic pipeline of multiplex imaging from raw images to statistical analysis and modeling. 

Cell phenotyping is the process by which the cells are labeled with a type. There are two main strategies for cell phenotyping, marker gating, and unsupervised clustering methods. Marker gating refers to the process by which the distributions of individual markers are visualized and a threshold is drawn to define cells that are positive and negative for the marker. Recently, a semi-automated method for marker gating specifically for MI data was proposed by \citet{xiong2023gammagater}. Unsupervised clustering methods, such as K-means, require manual annotations after clustering is performed \citep{kiselev2017sc3,kiselev2019challenges}. 


For each image, the phenotyped cells along with their spatial coordinates can be thought of as a multi-type point pattern. These point patterns are typically nested within a tissue section and tissue sections are nested within patients (see Figure \ref{fig:MI}). There may be additional covariates at each level of the hierarchy. For example, there may be functional markers measured at the cell level, tissue properties measured at the image level, and patient-level covariates. A formal statistical analysis of MI data is usually concerned with characterization and testing associations between first- and/or second-order properties of the point patterns and patient-level outcomes or covariates, as we detail below.  

\subsection{First-order analysis}\label{subsec:first}
The spatial locations of the cells and their respective cell-type labels can be thought of as a point pattern. Point patterns can be characterized by first- and second- order metrics.  First-order metrics measure the abundance of a certain cell type or the cell composition in an image. Examples of first-order metrics include, but are not limited to, the count, the proportion, or the spatial intensity of a cell type. Generalized linear models (GLMs) provide a framework for modeling count or proportions. For example, a Poisson GLM may be applied to quantify the association between the number of T cells in the TME and response to immunotherapy controlling for patient-level covariates such as cancer stage, age, and sex. The spatial intensity may be summarized as the number of cells per unit area and be modeled as a count or continuous variable. 
Alternatively, a logistic regression may be used to model proportions of T cells in conjunction with weights equal to the number of cells in each image.

The TME is known to be highly heterogeneous even within a single subject \citep{jca17648}. For this reason, several images may be collected for each subject. The GLM framework can be extended to account for correlations between images of the same subject by adding a subject-specific random effect \citep{McCullagh1989generalized}.
Generalized estimating equations (GEE) can be used to estimate the marginal associations while taking within-subject correlations into account \citep{liang1986longitudinal}. \citet{sanchez2021multiplex} used a Poisson mixed model to model the association between cell composition and time (pre vs post treatment) while accounting for correlated images within the same subject. 

Cell type counts or proportions in MI data are often highly over-dispersed, meaning that the variance of the observed data is greater than what is prescribed by the model. This can lead to inflated Type I error. A robust estimate of the standard error, known as the `sandwich' estimate, can be employed to account for possible overdispersion \citep{dean2014overdispersion}. Alternatively, the variance may be modeled with an additional scale parameter to account for overdispersion. In the GLM framework, these models are known as quasi-Poisson or quasi-Binomial \citep{sun2018heritability,gomez2020quasi}. Additional scale parameters may also be specified in GLMM or GEE models. 

Zero-inflation is also a common feature in MI data. This refers to the excess of zero counts, or images with no cells of a certain type. This can be common when measuring the proportion or count of a rare cell type. Zero-inflated Poisson (ZIP) or zero-inflated Negative Binomial (ZINB) models accommodate excess zeros by modeling the outcome with two components. When there are multiple images per subject, zero-inflated versions of GLMM and GEE may be utilized \citep{RJ-2017-066,sarvi2019gee}.

The outcome of interest may be survival or other time-to-event data. In these cases, cell composition can be aggregated for each patient and tested via Cox Regression or other time-to-event models \citep{patwa2021multiplexed, johnson_cancer_2021}.   The marker intensities may be directly used to assess and model cell composition in MI data. \citet{seal_denvar_2022} propose a distance-based clustering method using probability densities of the marker intensities. The group labels from the clustering algorithm may be tested for association in patient-outcome models.  

\subsection{Second-order analysis}\label{subsec:second}

A second-order metric of a point pattern quantifies the extent to which marks tend to attract or repel eachother in space. These metrics can be applied within cell types, measuring the extent to which one cell type tends to cluster with itself in space. Or, second-order metrics can be applied between pairs of cell types, measuring colocalization between types. There are many metrics which have been proposed to capture second-order properties. We enumerate several popular metrics below and note some of the statistical methods for use of these metrics in statistical modeling. 

\paragraph{Cross functions} Spatial summary metrics based on point process theory are popular for quantifying the extent to which cell types co-localize (or repel other cell types) in an image. These are typically based on different varieties of `cross' functions such as K- and G-cross functions and their variants. For example, one such metric is Ripley’s K \citep{ripley_second-order_1976}. For any pair of types $i$ and $j$, the multitype K-function $K_{ij}(r)$ (also known as the K-cross function), is the expected number of points of type $j$ lying within a distance $r$ or a typical point of type $i$, standardized by the intensity of points of type $j$,  $\lambda_j$ \citep{baddeley_spatial_2015}.  


The K-cross function can be estimated by the empirical K-cross function: 
\begin{equation}
    \hat{K}_{ij}(r) = \frac{|W|}{n_i n_j}\sum_{i=1}^{n_i}\sum_{j=1, j\ne i}^{n_j}\mathbbm{1}\{d_{ij}\leq r\}e_{ij}(r), \nonumber
\end{equation}
where $|W|$ is the area of the image, $n_j$ is the number of points of type $j$,  $n_i$ is the number of points of type $i$, $d_{ij}$ is the Euclidean distance between a point of type $i$ and a point of type $j$, and $e_{ij}(r)$ is an edge correction weight which accounts for bias due to unobserved points outside the window. Edge correction methods include border correction, isotropic correction, and translation correction.  The K-cross function can be compared to the theoretical K-function under complete spatial randomness (CSR) or independence between the points of type $i$ and $j$ in order to quantify the amount of clustering or anti-clustering between a pair of cell types. The theoretical version of the K-cross function is:  $K_{\text{theo}}(r)= \pi r^2$ which holds under assumed stationarity of the point pattern (see below for a discussion of stationarity). 

Other common summary metrics for spatial clustering are Besag's L-cross function and the cross-type pair correlation function--both based on the K-cross function and given by 
\begin{eqnarray}
    L_{ij}(r)=\sqrt{\frac{K_{ij}(r)}{\pi}} & \text{ and } g_{ij}(r)=\frac{K'_{ij}(r)}{2\pi r}, \nonumber
\end{eqnarray}
where $K'$ is the first derivative of the K function. The cross-type pair correlation function $g(r)$ is the probability of observing a pair of points separated by a distance of $r$, whereas the K-cross function is cumulative in nature. 

The multi-type G function (G-cross) between types $i$ and $j$ is based on the distance from a typical point of type $i$ to the nearest point of type $j$. The uncorrected empirical estimate of $G_{ij}(r)$ is given by
\begin{eqnarray}
    \hat{G}_{ij}(r)=\frac{1}{n_i}\sum_i\mathbbm{1}\{d(x_i^{(i)},{\bf x}^{(j)} \backslash x_i^{(i)}) \leq r\}, \nonumber
\end{eqnarray}
where $d(x_i^{(i)},{\bf x}^{(j)} \backslash x_i)$ is the distance between the $i$-th point of type $i$ and the nearest point of type $j$. This estimate can be biased for the true $G_{ij}(r)$ due to edge effects. Several edge correction methods exist for the G-cross function including, border correction, Kaplan-Meier correction, and Hanisch-Chiu-Stoyan correction \citep{Baddeley2019SpatialSA}. The G-cross function can be compared to the theoretical version of the G-function under CSR assuming stationarity, $G_{ij \text{theo}}=1-\exp(\lambda_j\pi r^2)$. Visualizations and comparisons between different summary functions including the ones discussed here are presented by \citet{parra2021methods}.  

While informative, the K-, L-, g-, and G- functions are sensitive to deviations from the assumed stationarity of the point process. Any holes or varying density of cells across the image are violations of stationarity. These violations often cause false positive indication of clustering between cells (i.e. the theoretical version of the function no longer represents how the function would behave under CSR). To address these issues, and ensure that the summary functions can be comparable between different point patterns, the observed K-, L-, g-, and G- functions can be compared to their respective functions under CSR using permutation envelopes. The envelop method begins by randomly permuting cell type labels to generate several `random' multi-type point patterns (each with the same underlying cell positions). The summary functions are calculated for each permuted point pattern, generating an envelop of how the function behaves in the presence of no clustering or anti-clustering between cell types. Then the observed function is compared to the CSR envelope to determine if there is clustering or anti-clustering over and above what could be expected due to the underlying locations of cells \citep{baddeley_spatial_2016}.  

Several state-of-the-art methods have been proposed recently to quantify the association between spatial summary functions such as K, L, g, and G at the image level with patient-level outcomes or covariates. \citet{canete_spicyr_2022} modeled the difference between the observed L-cross function and the theoretical L-cross function at several radii as the dependent variable in a linear model. Patient-specific random effects account for correlation between images of the same subject. A weighting scheme is applied to images proportionally to the number of cells they contain. \citet{seal_spaceanova_2023} use a functional analysis of variance approach to test differential spatial colocalization as quantified by the g-function across multiple tissue or disease groups. This method allows for multiple images per subject and utilizes permutations to account for missing tissue regions or inhomogeneity of the point patterns. \citet{vu_spf_2022} proposed a functional data approach to quantify the association between patient survival as the dependent variable and the mark connection function (a function of the g-function) as the independent variable. They model the mark connection function using a spline basis. The spatialTIME \citep{creed_spatialtime_2021} tool offers spatial analysis capabilities of MI data in the context of TME, including data preprocessing and clustering measures, to provide insights into the spatial heterogeneity of immune cells within tissue samples, aiding in the understanding of cancer biology.   
 
\paragraph{Density-based methods} Alternatives to the K, g, L, and G functions attempting to quantify cell colocalization have been recently proposed for MI data. \citet{krishnan_gawrdenmap_2022} used geographically weighted regression on smoothed intensity surfaces to quantify the amount of colocalization between cell type pairs. Then the densities of the geographically weighted regression coefficients are used to classify patients into distinct groups. More recently, \citet{masotti2023dimple} proposed a method called DIMPLE (distance matrices for multiplex imaging) which uses smoothed intensity estimates of the cell type locations and compares each pair of cell types using Jensen-Shannon distance. DIMPLE is computationally efficient and less sensitive to inhomogeneity of the point pattern than traditional point process methods. The distances are computed at the level of each image, thus requiring no additional summarizations. These distances can be readily compared within and across images in the data using statistical tests or models. The method can be implemented using the {\small DIMPLE} R package\footnote{\url{https://github.com/nateosher/DIMPLE}} along with a Shiny application\footnote{\url{https://bayesrx.shinyapps.io/DIMPLE_Shiny/}}.

\paragraph{Other methods} \citet{vu_funspace_2022} applied spatial entropy measures to quantify the spatial heterogeneity of multiple cell types simultaneously. They study how the variability of the TME is associated with patient-level outcomes. Several new statistical methods bypass the need for phenotyping cells entirely by modeling the marker intensities directly. \citet{chervoneva2021quantification} assume a marked point pattern where the marks are the continuous marker intensities. They proposed metrics based on the conditional mean and conditional variance of a mark given that there is another point of the process a distance away. Area under the curve is used to summarize the conditional mean and variance functions which can then be tested as covariates in a patient-outcome model. \citet{arnol_modeling_2019} used a random effect model framework to model protein marker intensity as a function of additive components of cell-state effects, location or environment effects, and cell-cell interaction effects. 

\subsection{Design considerations} The design of MI studies involves several decisions that require statistical thought. The TME is known to be highly heterogeneous even within one subject. With a fixed amount of resources and a required amount of statistical power and allowable Type I error, an investigator must decide how many images and patients are necessary. \citet{baker2023silico} simulate MI data and conduct a power analysis using first- and second-order spatial biomarkers. In the simulations, they vary the tissue size, diversity of cell types, spatial structure, and sampling strategy. \citet{bost2023optimizing} use spatial transcriptomic data to infer optimal tissue sampling strategy to identify all cell phenotypes of interest across various tissue samples. 


\section{Conclusions and Future Directions }\label{sec:discussion}

In this review article, we have attempted to summarize the key contributions that statisticians and modelers have made in quantitative medical imaging analyses. We have focused our review on radiology and pathology imaging, which are critical modalities for diagnosis, treatment, and prognosis of many diseases, especially cancer. We underscore the integration of sophisticated imaging and rigorous statistical techniques to enhance the understanding of tumor behaviours and predictive capabilities it can offer for answering various biological and clinical questions. Although technological and computational efforts have enabled rapid development of the field, many challenges and open problems remain -- both broadly and technology specific -- we expound on a few below. 

\paragraph{Artificial intelligence and data integration} One significant area of ongoing research is the development of more advanced methodologies and algorithms for handling the vast amounts of richly structured data generated by modern imaging technologies. A potential avenue would be hybridization of artificial intelligence (AI) and machine learning tools such as deep learning models and core statistical models that can be used to effectively manage, analyze, and more importantly interpret these complex, high-dimensional datasets. There is also a critical need for robust methods to integrate multimodal imaging data, which would allow for a more comprehensive understanding of disease mechanisms by correlating information from different imaging sources. Another open problem is improving the accuracy and reliability of imaging biomarkers in clinical settings, particularly in their ability to predict disease progression and response to therapy. Furthermore, addressing the computational demands and ensuring the privacy and security of sensitive medical imaging data remains paramount, especially as cloud-based and AI-driven technologies become increasingly prevalent. 
 

\paragraph{Radiology imaging}  In radiology imaging research, a significant challenge lies in image data harmonization, for example, when images are collected from different sites or scanners. While this issue has a lesser impact on shape analysis, it affects texture analysis due to the necessity of intensity normalization. Existing methods addressing this problem are disease or application specific. Consequently, there is a need for approaches that can establish general guidelines for image data harmonization, effectively leveraging the inherent structure of the data, as this directly influences downstream inference. Segmentation of regions of interest (e.g., tumor), a vital task in biomedical imaging analysis, has predominantly been addressed using machine learning-based methods, leaving ample space for statistical models. However, efficiency in computation is paramount given the high-dimensional nature of imaging data. Radiology imaging provides three-dimensional MRI scans, yet many methods designed to incorporate spatial context operate on two-dimensional slices chosen based on specific criteria. While volume and some texture-based features can be computed for three-dimensional datasets, methods for three-dimensional shape assessment are lacking, highlighting an open avenue for research to develop rigorous statistical models for shape and texture analysis. Existing methods summarize pertinent aspects from regions of interest in the image into a Euclidean representation, diminishing the interpretability of features and complicating direct linkage to biological context. Hence, there is a critical need for methods that retain interpretability while accommodating shape- or function-based features. Integrating radiology data with data from other platforms, including other imaging modalities, holds promise for providing crucial insights into disease progression. Consequently, methods facilitating such integration would significantly advance the field. Though longitudinal analysis is less common due to complexities in data acquisition, particularly for certain types of radiology imaging, there is a growing need for statistical modeling approaches tailored to longitudinal imaging data. In summary, the wide array of challenges and opportunities in the statistical analysis of radiology images underscores the immense scope for further research in the field.

\paragraph{Pathology imaging}  With respect to H\&E imaging analysis, it bears mentioning that there is a distinct pattern across the various methods presented. Regardless of the method used, the level of interaction on a given biopsy is ultimately summarized at the level of each biopsy before being used in downstream modeling. To the best of our knowledge, there has not been any explicit work focused on \textit{jointly} modeling separate whole slide H\&E images. This is likely the case for two primary reasons. First, this data lends itself most naturally to being modeled as a point process. Because most point process models operate on a single (though not necessarily contiguous) space, the general strategy of summarizing the modeling results at the biopsy level and using them in downstream analysis is the most natural way of accomplishing such joint modeling. Second, the spatial characteristics of tumors and tumor tissue samples are quite complicated. Often times, modeling across disparate spaces relies on leveraging a fundamental similarity of the spaces. For example, in neuroimaging analyses, population level modeling is possible because of the fundamental similarity in structure of the human brain across subjects. Unfortunately, neither tumors nor tissue samples from tumors share any remotely common spatial structure. Thus, the prospect of building joint spatial models across disparate H\&E images is distinctly challenging and an open problem.

\paragraph{Spatial multiplex imaging}  In terms of MI data, a common pitfall in such studies is that the first- and second-order relationships among cells in the TME rely on first obtaining the correct cell phenotypes. The uncertainty related to the preprocessing and phenotyping steps in the MI pipeline is often ignored in downstream analysis and modeling. Besides methods that operate directly on the marker intensities, there are no methods to our knowledge which integrate uncertainty in the cell phenotypes with downstream analyses. Another pitfall in MI studies is the potential amount of within patient heterogeneity. Large scale MI studies may only generate one or two images representative of a patient, it is an open question as to whether that is enough data to fully characterize a patient's TME. Furthermore, there are several other open questions in the field of spatial multiplex imaging data with respect to statistical approaches. First, there is a lack of statistical methods which attempt to quantify spatial interactions beyond three or more cell types at a time. These higher order interactions may signal unique cellular niches or neighborhoods. Second, as technologies rapidly improve, computational efficiency and scalability will be increasingly important. Any statistical approaches will need to scale with increasing numbers of images and increasing numbers of unique cell types. Finally, there may be protein markers or spatially varying features generated alongside the phenotypic markers, such as functional markers or tissue types. These spatially varying features may mediate spatial relationships between cell types. Quantification of this phenomenon via statistical models will be increasingly important as technologies improve allowing for collection of more and more markers. 
Furthermore, cancer progression is driven by an agglomeration of genetic mutations which unfolds spatially and the process involves the disruption of normal tissue structure, invasion into surrounding areas, and eventual metastasis. Tumors are embedded within a dynamic ecosystem comprising both the cancer cells and their surrounding microenvironment. Modern spatial genomic, transcriptomic, and proteomic technologies now provide unprecedented insights into this intricate interplay \cite{Seferbekova2023}. Development of integrative spatial methods that harness these sequencing-based genomic, transcriptomic, and proteomic spatial layers along with imaging-based data can provide a more comprehensive characterization of tumor microenvironment.








\section*{Acknowledgements and Disclosure Statement}
This work was supported by NCI grants, P30 CA046592 (M.M. and V.B.), R37 CA214955-01A1 (M.M. and V.B.), R01 CA244845-01A1 (V.B.), 5T32CA083654 (V.B. and N.O.), Michigan Institute for Data Science grant (M.M.), Rogel Cancer Center funds (N.O., S.A. and V.B.), NIH grant U19-AG068753 (S.M.), NINDS grant R21 NS135268-01 (S.M.), NIGMS grant R01 GM127430-06A1 (S.M.), Boston University Population Health Data Science Seed Funds (S.M.) and Institutional Research Funds (S.M.). The authors declare no competing interests. 





\newpage
\bibliography{bibliography}
\end{document}